\documentclass[aps,twocolumn,a4paper,nofootinbib,superscriptaddress]{revtex4}
\usepackage[dvips]{graphicx}
\usepackage{amsmath}
\usepackage{amssymb}

\begin{document}

\title{Many-body filling-factor dependent renormalization of Fermi velocity in graphene\\ in strong magnetic field}
\author{Alexey A. Sokolik}%
\affiliation{Institute for Spectroscopy, Russian Academy of Sciences, 142190 Troitsk, Moscow, Russia}%
\affiliation{National Research University Higher School of Economics, 109028 Moscow, Russia}%
\author{Yurii E. Lozovik}%
\email{lozovik@isan.troitsk.ru}
\affiliation{Institute for Spectroscopy, Russian Academy of Sciences, 142190 Troitsk, Moscow, Russia}%
\affiliation{National Research University Higher School of Economics, 109028 Moscow, Russia}
\affiliation{Dukhov Research Institute of Automatics (VNIIA), 127055 Moscow, Russia}%

\begin{abstract}
We present the theory of many-body corrections to cyclotron transition energies in graphene in strong magnetic field
due to Coulomb interaction, considered in terms of the renormalized Fermi velocity. A particular emphasis is made on
the recent experiments where detailed dependencies of this velocity on the Landau level filling factor for individual
transitions were measured. Taking into account the many-body exchange, excitonic corrections and interaction screening
in the static random-phase approximation, we successfully explained the main features of the experimental data, in
particular that the Fermi velocities have plateaus when the 0th Landau level is partially filled and rapidly decrease
at higher carrier densities due to enhancement of the screening. We also explained the features of the nonmonotonous
filling-factor dependence of the Fermi velocity observed in the earlier cyclotron resonance experiment with disordered
graphene by taking into account the disorder-induced Landau level broadening.
\end{abstract}

\maketitle

\section{Introduction}
Massless Dirac electrons in single-layer graphene offer an opportunity to study condensed-matter counterparts of
relativistic effects and to achieve new regimes in quantum many-body systems \cite{CastroNeto,Novoselov,Kotov}.
Low-energy electronic excitations in this material obey the Dirac equation and move with the constant Fermi velocity
$v_\mathrm{F}\approx10^6\,\mbox{m/s}$. In a strong perpendicular magnetic field $B$, quantization of an electron
kinetic energy in graphene results in the relativistic Landau levels \cite{Goerbig}
\begin{eqnarray}
E_n=\mathrm{sgn}(n)\,v_\mathrm{F}\sqrt{2|n|Be\hbar/c},\quad n=0,\pm1,\pm2,\ldots\label{E_n}
\end{eqnarray}
Unlike usual Landau levels for massive electrons, the relativistic ones are not equidistant $E_n\propto\sqrt{|n|}$,
scale as a square root of magnetic field, $E_n\propto\sqrt{B}$, and obey the electron-hole symmetry, $E_n=-E_{-n}$.
Relativistic nature of graphene Landau levels was first confirmed by the half-integer quantum Hall effect
\cite{Novoselov}, and direct observations of these levels using the scanning tunneling spectroscopy had followed (see
the review of experiments in \cite{Yin}).

Another way to study Landau levels in graphene is to induce electron interlevel transitions by an electromagnetic
radiation, typically in the infrared range. The selection rules for photon absorption \cite{Sadowski1} require
$\Delta|n|=\pm1$, implying the intraband $-n-1\rightarrow-n$, $n\rightarrow n+1$, and interband $-n-1\rightarrow n$
(which will be referred to as $\mathrm{T}_{n+1}^-$) and $-n\rightarrow n+1$ (referred to as $\mathrm{T}_{n+1}^+$)
transitions. The interband transitions $\mathrm{T}_{n+1}^\pm$, which are more widely studied, have the energies
\begin{eqnarray}
E_{n+1}-E_{-n}=v_\mathrm{F}\sqrt{2Be\hbar/c}\left(\sqrt{n}+\sqrt{n+1}\right)\label{Omega_n_0}
\end{eqnarray}
in the ideal picture of massless Dirac electrons (\ref{E_n}) in the absence of interaction and disorder.

In a series of cyclotron resonance measurements, mainly on epitaxial graphene, transition energies in very good
agreement with Eq.~(\ref{Omega_n_0}) were reported (see \cite{Orlita,Basov} and references therein). However, the other
experiments \cite{Jiang1,Jiang2,Henriksen,Russell} demonstrated deviations from Eq.~(\ref{Omega_n_0}) due to many-body
effects and, possibly, disorder. Similar deviations were discovered in magneto-Raman scattering for both cyclotron
$\mathrm{T}_{n+1}^\pm$ \cite{Sonntag} and symmetric interband $-n\rightarrow n$ \cite{Berciaud,Faugeras1,Faugeras2}
transitions. Indeed, the Kohn's theorem \cite{Kohn}, which protects cyclotron resonance energies of usual massive
electrons against many-body renormalizations, is not applicable to graphene
\cite{Iyengar,Bychkov,Barlas,Roldan1,Shizuya1,Shizuya2,Gorbar,Chizhova,Menezes,Shizuya4,Sokolik1,Shizuya2018}. The
observed energies of $\mathrm{T}_{n+1}^\pm$ can be described by the counterpart of Eq.~(\ref{Omega_n_0})
\begin{eqnarray}
\Omega_{n+1}^\pm=v_\mathrm{F}^*\sqrt{2Be\hbar/c}\left(\sqrt{n}+\sqrt{n+1}\right)\label{Omega_n}
\end{eqnarray}
with the bare Fermi velocity $v_\mathrm{F}$ replaced by the renormalized velocity $v_\mathrm{F}^*$. While the former
one, $v_\mathrm{F}$, should be close to $0.85\times10^6\,\mbox{m/s}$, as indicated by fitting theoretical calculations
to various experimental data on graphene (see, e.g., \cite{Sokolik1,Yu,Elias,YangPRL}), the latter one,
$v_\mathrm{F}^*$, range from $10^6\,\mbox{m/s}$ to $1.4\times10^6\,\mbox{m/s}$ depending on carrier density, magnetic
field and substrate material \cite{Jiang1,Jiang2,Henriksen,Russell,Berciaud,Faugeras1,Faugeras2,Sonntag}. The existing
theory describes renormalization of Fermi velocity in magnetic field with reasonable accuracy in the Hartree-Fock
\cite{Shizuya1,Shizuya2,Faugeras1,Chizhova} and static random-phase \cite{Gorbar,Sokolik1,Shovkovy} approximations.

In two very recent experiments \cite{Sonntag,Russell}, the energies of the $\mathrm{T}_{n+1}^\pm$ transitions were
measured with high accuracy as functions of the Landau level filling factor $\nu$, that may provide an especially deep
insight into the many-body physics of graphene in magnetic field. Unlike graphene without magnetic field, where
$v_\mathrm{F}^*$ diverges logarithmically upon approach to the charge neutrality point \cite{Kotov,Gonzalez,Elias},
here it saturates to a finite value at $\nu\rightarrow0$, and, in the most cases, has even a broad plateau in the range
$-2<\nu<2$.

In this article, we calculate the energies of the $\mathrm{T}_{n+1}^\pm$ transitions as functions of the filling factor
$\nu$ with taking into account many-body effects. Our approach, which is described in Sec.~\ref{sec2} and
Appendices~\ref{Appendix_A}, \ref{Appendix_B}, and \ref{Appendix_C}, takes into account the screening of the Coulomb
interaction as one of the key points, that contrasts with the most calculations on this subject
\cite{Iyengar,Bychkov,Barlas,Roldan1,Shizuya1,Shizuya2,Faugeras1,Chizhova,Shizuya4,Shizuya2018} based on the
Hartree-Fock approximation with unscreened interaction. The screening allowed us to describe experimental data on both
Landau levels \cite{Sokolik2} and interlevel transition energies \cite{Sokolik1} earlier, and provides improved
understanding to the filling-factor dependence of the observed $v_\mathrm{F}^*$ in this work as well.

In Sec.~\ref{sec3} we analyze the electron-hole asymmetry of transition energies and the presence of plateaus at
$-2<\nu<2$, following from the properties of interaction matrix elements. In Sec.~\ref{sec4} we present the results of
numerical calculations, which reproduce the main features of the experimental the $v_\mathrm{F}(\nu)$ dependencies from
Refs.~\cite{Sonntag,Russell}: a) the plateaus in $v_\mathrm{F}^*$ at $-2<\nu<2$ when the 0th Landau level is partially
filled, b) the rapid decrease of $v_\mathrm{F}^*$ at $|\nu|>2$ with increasing carrier density, c) the decrease of
$v_\mathrm{F}^*$ at $\nu=\mathrm{const}$ at increasing magnetic field. We have found good agreement between the
experiments and the theory using the bare Fermi velocity $v_\mathrm{F}=0.85\times10^6\,\mbox{m/s}$ and realistic values
of the dielectric constant $\varepsilon$. The intraband transitions $n\rightarrow n+1$ and $-n-1\rightarrow n$ were
also analyzed, and we predict the V-shaped dependence of their energies on $\nu$.

Additionally, we have considered the nonmonotonous dependence $v_\mathrm{F}^*(\nu)$ for the $\mathrm{T}_1^\pm$
transition observed in \cite{Henriksen} with the maximum at $\nu=0$ and minima at $\nu=\pm2$. Taking into account a
disorder-induced broadening of Landau levels, we have explained this dependence with good accuracy in Sec.~\ref{sec5}.
Our conclusions are presented in Sec.~\ref{sec6}.

\section{Theoretical approach}\label{sec2}

Dynamical conductivity of graphene can be calculated using the Kubo formula \cite{Gusynin}
\begin{eqnarray}
\sigma_{\alpha\beta}(\mathbf{q},\omega)=\frac1{\hbar\omega S}\int\limits_0^\infty
dt\:e^{i(\omega+i\delta)t}\langle[j_\alpha(\mathbf{q},t),j_\beta(-\mathbf{q},0)]\rangle,\label{sigma1}
\end{eqnarray}
where $j_\alpha(\mathbf{q},t)$ is the $\alpha$-axis projection of the Fourier component of the current density operator
$j_\alpha(\mathbf{q})=ev_\mathrm{F}\int d\mathbf{r}\:\Psi^+(\mathbf{r})\sigma_\alpha\Psi(\mathbf{r})e^{-i\mathbf{qr}}$
evolving in time in the Heisenberg representation, $\Psi(\mathbf{r})$ is the two-component field operator for Dirac
electrons, $S$ is the system area, and $\delta\rightarrow+0$.

\begin{figure}[t]
\begin{center}
\resizebox{0.9\columnwidth}{!}{\includegraphics{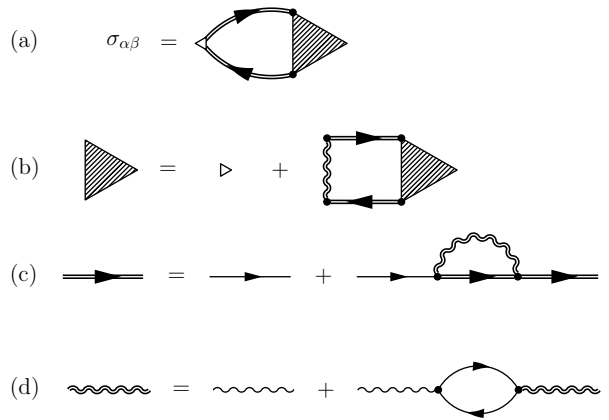}}
\end{center}
\caption{\label{Fig1} (a) Diagrammatic relationship (\ref{Gj}) between the current Green function and the vertex. (b),
(c) Equations for, respectively, the vertex function and the electron Green function in the mean-field approximation.
(d) Coulomb interaction screening in the random-phase approximation.}
\end{figure}

Diagrammatic representation of the conductivity, shown in Fig.~\ref{Fig1}(a), allows its calculation in terms of the
current vertex matrix $\Gamma_\beta$, which would be equal to $\sigma_\beta$ in the absence of interaction and
disorder. To find it, we use the mean field approximation, where the excitonic ladder [Fig.~\ref{Fig1}(b)] for the
vertex $\Gamma_\beta$ and the one-loop self-energy corrections [Fig.~\ref{Fig1}(c)] for the single-particle Green
functions $G$ are taken into account. Using the interaction, which is statically screened in the random-phase
approximation [Fig.~\ref{Fig1}(d)], greatly simplify the calculations. If we additionally neglect the mixing of
different pairs of electron and hole Landau levels, appearing in the excitonic ladder (which was shown to be weak under
typical conditions with using the screened interaction \cite{Sokolik1}), the optical conductivity
$\sigma_{\alpha\beta}(\omega)\equiv\sigma_{\alpha\beta}(0,\omega)$ is (see the details of calculations in
Appendix~\ref{Appendix_C}):
\begin{eqnarray}
\sigma_{\alpha\beta}(\omega)=\frac{ie^2v_\mathrm{F}^2}\omega\sum_{n_1n_2}
\frac{f_{n_2}-f_{n_1}}{\hbar\omega-\Omega_{n_1n_2}+i\delta}\nonumber\\
\times\mathrm{Tr}\left[\Phi_{n_1n_2}(0)\sigma_\alpha\right]
\mathrm{Tr}\left[\Phi^+_{n_1n_2}(0)\sigma_\beta\right].\label{sigma3}
\end{eqnarray}
Here $f_n$ is the occupation number ($0\leqslant f_n\leqslant 1$) of the $n$th Landau level, and the matrix
$\Phi_{n_1n_2}(0)$, which is defined by (\ref{Phi_nk}) and (\ref{phi_nk}), determines the selection rules
$|n_1|=|n_2|\pm1$ for each $n_2\rightarrow n_1$ transition. The resonant transition energy $\Omega_{n_1n_2}$, where
$\sigma_{\alpha\beta}$ has a pole, consists of the difference between the bare Landau level energies $E_{n_1}-E_{n_2}$,
the difference between electron self-energies $\Sigma_{n_1}-\Sigma_{n_2}$, and the excitonic correction $\Delta
E_{n_1,n_2}^\mathrm{(exc)}$ (see the similar formula in \cite{Shizuya2018}):
\begin{eqnarray}
\Omega_{n_1n_2}=E_{n_1}-E_{n_2}+\Sigma_{n_1}-\Sigma_{n_2}+\Delta E_{n_1n_2}^\mathrm{(exc)}.\label{Omega}
\end{eqnarray}

In the mean field approximation, the self-energy
\begin{eqnarray}
\Sigma_n=-\sum_{n'}f_{n'}\langle nn'|V|n'n\rangle,\label{Sigma_exch}
\end{eqnarray}
as shown in Appendix~\ref{Appendix_B}, is a sum of the exchange matrix elements
\begin{eqnarray}
\langle nn'|V|n'n\rangle=2^{\delta_{n0}+\delta_{n'0}-2}\frac{l_H^2}{2\pi}\int d\mathbf{q}\:V(q)\nonumber\\
\times\left|s_ns_{n'}\phi_{|n|-1,|n'|-1}(\mathbf{a}_\mathbf{q}) +\phi_{|n||n'|}(\mathbf{a}_\mathbf{q})\right|^2
\label{exch_me}
\end{eqnarray}
of the screened Coulomb interaction $V(q)$ between the $n$th and all filled $n'$th Landau levels
\cite{Iyengar,Bychkov,Faugeras1}, where the functions $\phi_{nk}$ are defined in (\ref{phi_nk}), and
$\mathbf{a}_\mathbf{q}\equiv-l_H^2[\mathbf{e}_z\times\mathbf{q}]$.

The excitonic correction
\begin{eqnarray}
\Delta E_{n_1n_2}^\mathrm{(exc)}=-(f_{n_2}-f_{n_1})\langle n_1n_2|V|n_1n_2\rangle\label{E_exc}
\end{eqnarray}
is the direct interaction matrix element
\begin{eqnarray}
\langle n_1n_2|V|n_1n_2\rangle=2^{\delta_{n_10}+\delta_{n_20}-2}\frac{l_H^2}{2\pi}\int d\mathbf{q}\:V(q)\nonumber\\
\times\left\{\phi^*_{|n_1|-1,|n_1|-1}(\mathbf{a}_\mathbf{q}) +\phi^*_{|n_1||n_1|}(\mathbf{a}_\mathbf{q})\right\}
\label{exc_me}\\
\times\left\{\phi_{|n_2|-1,|n_2|-1}(\mathbf{a}_\mathbf{q}) +\phi_{|n_2||n_2|}(\mathbf{a}_\mathbf{q})\right\}\nonumber
\end{eqnarray}
with the minus sign, weighted with the difference of occupation numbers of the final and initial levels.

The dynamically screened interaction in the random-phase approximation is [see Fig.~\ref{Fig1}(d)]
\begin{eqnarray}
V(q,i\omega)=\frac{v_q}{1-v_q\Pi(q,i\omega)},\label{V_scr}
\end{eqnarray}
where $v_q=2\pi e^2/\varepsilon q$ is the bare Coulomb interaction weakened by the surrounding medium with the
dielectric constant $\varepsilon$, and
\begin{eqnarray}
\Pi(q,i\omega)=g\sum_{nn'}F_{nn'}(q)\frac{f_n-f_{n'}}{i\omega+E_n-E_{n'}},\label{Pi}
\end{eqnarray}
is the polarizability (or density response function) of noninteracting Dirac electrons
\cite{Goerbig,Roldan2,Lozovik,Roldan3,Pyatkovskiy,Gumbs}. Here
\begin{eqnarray}
F_{nn'}(q)=2^{\delta_{n0}+\delta_{n'0}-2}\nonumber\\
\times\left|s_ns_{n'}\phi_{|n|-1,|n'|-1}(\mathbf{a}_\mathbf{q})+\phi_{|n||n'|}(\mathbf{a}_\mathbf{q})\right|^2
\label{F_nnp}
\end{eqnarray}
is the form-factor of Landau level wave functions and $g=4$ is the degeneracy of electron states by valleys and spin.
The statically screened interaction $V(q)$ is obtained from (\ref{V_scr}), (\ref{Pi}) by taking $i\omega=0$.

In our model, there are three mechanisms leading to dependence of $\Omega_{n_1n_2}$ on the filling factor $\nu$ via the
occupation numbers
\begin{eqnarray}
f_n=\left\{\begin{array}{lll}0,&\mbox{if}&\nu\leq4n-2,\\
(\nu-4n+2)/4,&\mbox{if}&4n-2<\nu<4n+2,\\1,&\mbox{if}&\nu\geq4n+2,\end{array}\right.\label{f_n}
\end{eqnarray}
i.e.: through exchange energies (\ref{Sigma_exch}), excitonic corrections (\ref{E_exc}), and polarizability (\ref{Pi}).

Note that the sum (\ref{Sigma_exch}) over the filled Landau levels $n'$ in the valence band diverges at
$n'\rightarrow-\infty$, so we impose the cutoff $n'\geqslant-n_\mathrm{c}$ to obtain finite results. The physical
reason of thus cutoff is a finite actual number of Landau levels in the valence band, which can be found from the total
electron density: $n_\mathrm{c}=2\pi\hbar c/\sqrt3a^2eH\approx39600/B[\mbox{T}]$, where $a\approx2.46\,\mbox{\AA}$
\cite{Sokolik1,Sokolik2}.

\begin{figure}[t]
\begin{center}
\resizebox{1.0\columnwidth}{!}{\includegraphics{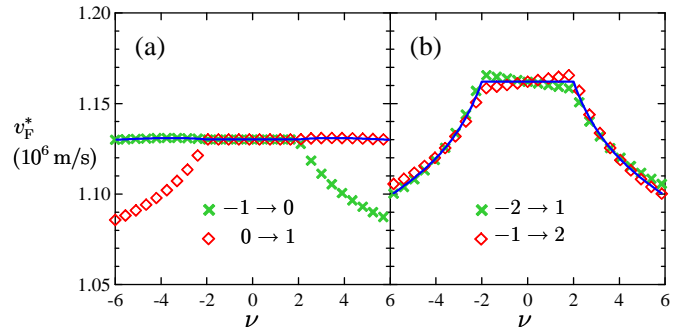}}
\end{center}
\caption{\label{Fig2} Renormalized Fermi velocities $v_\mathrm{F}^*$ for (a) the $T_1^\pm$ and (b) the $T_2^\pm$
transitions calculated with the screened interaction at $v_\mathrm{F}=0.85\times10^6\,\mbox{m/s}$, $\varepsilon=3.27$,
$B=8\,\mbox{T}$. Solid lines show the velocities found from the weighted transition energies (\ref{Omega_av}).}
\end{figure}

\section{Electron-hole asymmetry and plateaus at $-2<\nu<2$}\label{sec3}
The selection rule $|n_1|=|n_2|\pm1$ for the interband $n_2\rightarrow n_1$ transitions implies $n_1,n_2=n+1,-n$ (the
$\mathrm{T}_{n+1}^+$ transition) or $n_1,n_2=n,-n-1$ (the $\mathrm{T}_{n+1}^-$ transition). In the idealized Dirac
model without interactions, the energies of these transitions (\ref{Omega_n_0}) are equal. However this is no longer
the case when exchange self-energies are taken into account. Any nonzero doping $\nu\neq0$ introduces an asymmetry
between $\Omega_{n+1}^+$ and $\Omega_{n+1}^-$, at least, in the mean-field approximation. Looking at (\ref{Omega}) and
taking into account that $\langle n+1,-n|V|n+1,-n\rangle=\langle-n-1,n|V|-n-1,n\rangle$, we have:
\begin{eqnarray}
\Omega_{n+1}^+-\Omega_{n+1}^-=\Sigma_{n+1}+\Sigma_{-n-1}-\Sigma_{n}-\Sigma_{-n}\nonumber\\
+(f_n+f_{n+1}-f_{-n}-f_{-n-1})\langle n+1,-n|V|n+1,-n\rangle.\label{asym}
\end{eqnarray}
The electron-hole asymmetry in graphene, which is induced by the exchange interaction in the absence of magnetic field
and is similar in scale to our case, was found in \cite{Kretinin}.

The first line of (\ref{asym}) is a contribution of exchange self-energies to the asymmetry. Let us separate the
occupation numbers $f_{n'}=f_{n'}^{(0)}+\Delta f_{n'}$ on those of undoped graphene $f_{n'}^{(0)}$ and the
doping-induced part $\Delta f_{n'}$, and define $\Sigma^{(0)}_n=-\sum_{n'}f_{n'}^{(0)}\langle nn'|V|n'n\rangle$. Using
(\ref{exch_me}) and (\ref{k_sum}), and  neglecting a difference of small matrix elements at $n'\approx-n_\mathrm{c}$,
we get $\Sigma^{(0)}_{n+1}+\Sigma^{(0)}_{-n-1}-\Sigma^{(0)}_{n}-\Sigma^{(0)}_{-n}=0$. Thus the exchange energy
contribution to (\ref{asym}) arises only at nonzero doping $\nu\neq0$.

The second line of (\ref{asym}) corresponding to excitonic effects is nonzero only when either $\pm n$th or
$\pm(n+1)$th level is partially filled, i.e. at $4n-2<|\nu|<4n+6$. Since the polarizability (\ref{Pi}) and hence the
screened interaction $V(q)$ are even functions of $\nu$, both parts of (\ref{asym}) change sign at
$\nu\rightarrow-\nu$, so
\begin{eqnarray}
\Omega_{n+1}^+(\nu)=\Omega_{n+1}^-(-\nu).\label{conj}
\end{eqnarray}
This is illustrated in Fig.~2, where the typical calculated $v_\mathrm{F}^*$ are shown as functions of $\nu$.

The case of $n=0$ is the special one. The explicit structure of the wave functions (\ref{phi_nk}) imply the following
relationships connecting the matrix elements of direct and exchange interaction (valid even for non-Coulomb
potentials):
\begin{eqnarray}
\langle\pm1,0|V|\pm1,0\rangle+\langle\pm1,0|V|0,\pm1\rangle=\langle00|V|00\rangle.
\end{eqnarray}
In result, the doping-induced changes of exchange and excitonic parts of (\ref{Omega}) due to $f_0$ cancel each other
at ${-2<\nu<2}$, when the 0th level is partially filled. Additionally, the polarizability (\ref{Pi}) and hence $V(q)$
are also unchanged in this range of $\nu$, thus we expect plateaus in both $\Omega_1^\pm$:
\begin{eqnarray}
\Omega_1^+(\nu)=\Omega_1^-(-\nu)=\mathrm{const}\quad\mbox{at}\quad-2<\nu<2,
\end{eqnarray}
as seen in Fig.~\ref{Fig2}(a). For $n\neq0$ this is no longer the case, although variations of $\Omega_{n+1}^\pm(\nu)$
at $-2<\nu<2$ are typically very small [see Fig.~\ref{Fig2}(b)].

In experiments, $\Omega_{n+1}^+$ and $\Omega_{n+1}^-$ can be separated by observing cyclotron resonant absorption of
light with opposite circular polarizations. Using linear polarization, one can observe a mixture of these transitions
with relative intensities $I_{n+1}^+=f_{-n}-f_{n+1}$ and $I_{n+1}^-=f_{n}-f_{-n-1}$, equal to occupation number
differences in final and initial states. Assuming that experimental apparatus does not resolve the individual lines
$\Omega_{n+1}^+$ and $\Omega_{n+1}^-$, we will calculate the weighted transition energy
\begin{eqnarray}
\langle\Omega_{n+1}\rangle=\frac{\Omega_{n+1}^+I_{n+1}^++\Omega_{n+1}^-I_{n+1}^-}{I_{n+1}^++I_{n+1}^-}
\end{eqnarray}
and compare it with the experiments in the next section. From the particle-hole symmetry relationship
$f_n(-\nu)=1-f_n(\nu)$ we see that $\langle\Omega_{n+1}\rangle$ is even function of $\nu$. At $-2<\nu<2$,
$\langle\Omega_{n+1}\rangle$ is linear (via $f_0$) and at the same time even function of $\nu$, so
\begin{eqnarray}
\langle\Omega_{n+1}\rangle=\mathrm{const}\quad\mbox{at}\quad-2<\nu<2,\label{Omega_av}
\end{eqnarray}
as seen in Figs.~\ref{Fig2}(a,b). Thus our model predicts plateaus in all weighted transition energies
$\langle\Omega_{n+1}\rangle$ at $-2<\nu<2$. Similar conclusions about existence of the electron-hole asymmetry
(\ref{asym}) and the conjugation property (\ref{conj}) were made in the recent theoretical work \cite{Shizuya2018},
which considers transition energies in the Hartree-Fock approximation.

\begin{figure*}[t]
\begin{center}
\resizebox{1\textwidth}{!}{\includegraphics{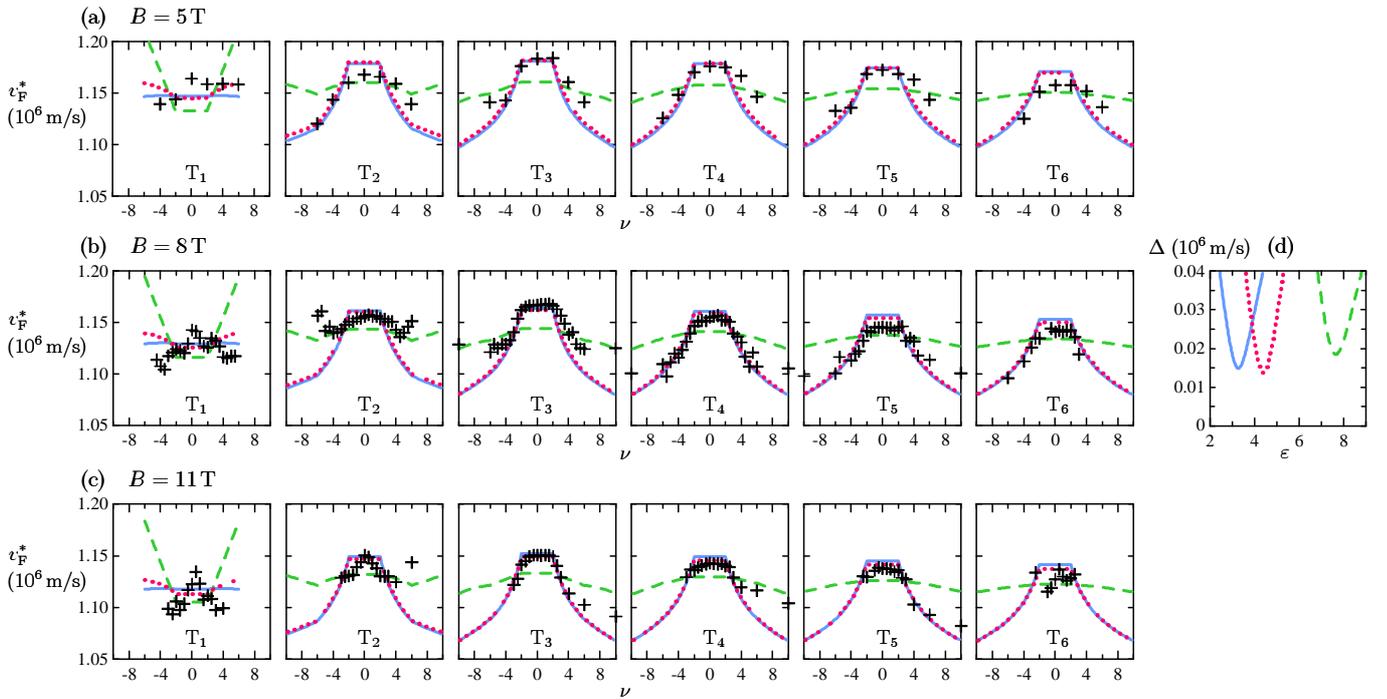}}
\end{center}
\caption{\label{Fig3} Renormalized Fermi velocities $v_\mathrm{F}^*$ at (a) $B=5$, (b) $8$, and (c) $11\,\mbox{T}$ for
the set of the $T_{n+1}$ transitions ($-n\rightarrow n+1/-n-1\rightarrow n$), taken from the experiment \cite{Russell}
(crosses), and calculated theoretically in the Hartree-Fock approximation (dashed lines), with taking into account the
interaction screening (solid lines) and with the self-consistent screening (dotted lines). The dielectric constants
$\varepsilon$, used in each calculation, are listed in the first line of Table~\ref{Table1}. Root mean square
deviations (\ref{rms}) between the calculated and experimental $v_\mathrm{F}^*$ are also shown (d) as functions of
$\varepsilon$ in the three approximations.}
\end{figure*}

\section{Calculation results}\label{sec4}
First we compare our calculations of the renormalized Fermi velocities
\begin{eqnarray}
v_\mathrm{F}^*=\frac{\langle\Omega_{n+1}\rangle}{\sqrt{2Be\hbar/c}\left(\sqrt{n}+\sqrt{n+1}\right)}
\end{eqnarray}
with the data of Ref.~\cite{Russell} where $\Omega_1\ldots\Omega_6$ as functions of $\nu$ were measured at three
magnetic fields $B=5,8,11\,\mbox{T}$. We fit the experimental points in three approximations:

1) Hartree-Fock approximation, where the unscreened Coulomb potential $v_q$ is used in all calculations.

2) Static random-phase approximation, where the potential $V(q)$ is screened (\ref{V_scr}) with using the
polarizability of noninteracting electron gas in magnetic field.

3) Self-consistent screening approximation, where the polarizability is multiplied by $v_\mathrm{F}/v_\mathrm{F}^*$ to
take into account weakening of the screening caused by many-body increase of the energy differences $E_{n'}-E_n$ in
denominators of (\ref{Pi}). Similarly to our previous works \cite{Sokolik1,Sokolik2}, this semi-phenomenological model
is aimed to achieve a self-consistency between many-body renormalizations of transition energies and screening. Using
the iterative procedure, we take $v_\mathrm{F}^*$, obtained on each step, to renormalize the screening when calculating
new $v_\mathrm{F}^*$ on the next step. About 5-6 iterations are usually sufficient to achieve a convergence.

Calculations in our approach depend only on two parameters: the bare Fermi velocity $v_\mathrm{F}$ and the dielectric
constant of a surrounding medium $\varepsilon$. In principle, both $v_\mathrm{F}$ and $\varepsilon$ can be treated as
fitting parameters. However variation of $v_\mathrm{F}$ in the range $(0.8\div0.95)\times10^6\,\mbox{m/s}$ allows to
achieve almost equally good agreement with the experimental data at slightly different $\varepsilon$, so a simultaneous
fitting of both parameters does not provide reliable results. Therefore we choose a specific value
$v_\mathrm{F}=0.85\times10^6\,\mbox{m/s}$, which was concluded to be the most suitable one based on theoretical fits of
several experimental data on graphene both in presence \cite{Sokolik1,Yu} and absence \cite{Elias,YangPRL} of magnetic
field. After that, the optimal dielectric constant of the surrounding medium $\varepsilon$ is the only adjustable
parameter in our model, and we find it by performing the least square fitting of the experimental points for all $n$
and $B$ simultaneously. Nevertheless it should be kept in mind that our fitting results can slightly change
quantitatively with different choice of $v_\mathrm{F}$ (although qualitative conclusions will be the same), and it
could be promising to implement a renormalization-group scheme for the Landau level data on graphene where all
unobservable variables like $v_\mathrm{F}$ can be excluded from the model.

\begin{table}[b]
\caption{\label{Table1}Dielectric constants of surrounding medium $\varepsilon$, which provide the best least-square
fittings of the experimental data from Refs.~\cite{Russell,Sonntag} at $v_\mathrm{F}=0.85\times10^6\,\mbox{m/s}$ in the
three approximations for the interaction listed in Sec.~\ref{sec4}.} \centering
\begin{tabular}{ccccc}
\hline%
\hline%
&Unscreened&Screened&Self-consistent\\
Experiment&interaction&interaction&screening\\
\hline%
Russell et al. \cite{Russell}&7.72&3.27&4.36\\
Sonntag et al. \cite{Sonntag}&5.50&1.05&2.55\\
\hline%
\hline%
\end{tabular}
\end{table}

The first line of Table~\ref{Table1} shows the optimal $\varepsilon$ used to fit the cyclotron resonance data of
Ref.~\cite{Russell} where high-mobility graphene samples were encapsulated from both sides in hexagonal boron nitride
monolayers and placed on an oxidized silicon. Fig.~\ref{Fig3} shows the experimental points together with our
calculations at these $\varepsilon$ in the three approximations described above. The calculation with the unscreened
Coulomb interaction (Hartree-Fock approximation) demonstrates two significant drawbacks. First, the dielectric constant
$\varepsilon\approx7.72$ is unrealistically high, because in this approximation it should imitate the interaction
screening by Dirac electrons in graphene in addition to the screening by an external medium. Second, the falloff of
$v_\mathrm{F}^*$ at $|\nu|>2$ turns out to be insufficient, because the increase of the screening strength (and,
consequently, suppression of the upward renormalization of the Fermi velocity) following the carrier density, is absent
here. For the $T_1$ transitions, the calculated $v_\mathrm{F}^*$ even increases at $|\nu|>2$, in contradiction with the
experiment, because the excitonic correction (\ref{E_exc}), which normally decrease $v_\mathrm{F}^*$, become suppressed
due to partial filling of $1$th or $-1$th Landau level. The similar drawbacks of the Hartree-Fock approximations were
mentioned in our previous works \cite{Sokolik1,Sokolik2}.

The screening allows us to achieve much better agreement with the experimental points at more realistic
$\varepsilon\approx3.27$, and the falloff of $v_\mathrm{F}^*$ at $|\nu|>2$ is reproduced very well. The iterative
calculations with the self-consistent screening provide almost the same curves, but at somewhat higher
$\varepsilon\approx4.36$. This distinction arises because the higher $\varepsilon$ is needed to compensate the
screening weakening caused by an upward renormalization of energy denominators in (\ref{Pi}).

\begin{figure*}[t]
\begin{center}
\resizebox{0.6\textwidth}{!}{\includegraphics{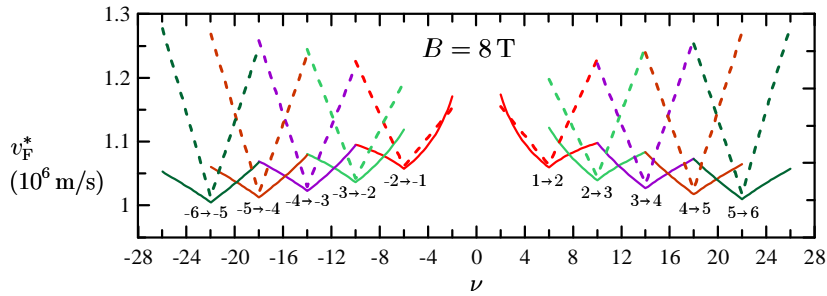}}
\end{center}
\caption{\label{Fig4} Renormalized Fermi velocities $v_\mathrm{F}^*$ for the intraband $n\rightarrow n+1$ and
$-n-1\rightarrow-n$ transitions calculated at $B=8\,\mbox{T}$ in the Hartree-Fock approximation (dashed lines) and with
taking into account the interaction screening (solid lines). The self-consistent iterative calculations are not shown
because their results are close to those with the non-self-consistent screening. The dielectric constants $\varepsilon$
are taken from the first line of Table~\ref{Table1}.}
\end{figure*}

Our calculations with taking into account the screening are thus able to fit the data of Ref.~\cite{Russell} at three
different $B$ and for six resonances $T_{n+1}$ simultaneously with the single adjustable parameter $\varepsilon$. We
can explain both the decrease of $v_\mathrm{F}^*$ at $\nu=0$ as $B$ gets higher, the plateaus at $|\nu|<2$, and the
rapid falloff of $v_\mathrm{F}^*$ at $|\nu|>2$, $n\geqslant1$ due to increase of the screening strength. The exceptions
are some inconsistencies of $v_\mathrm{F}^*$ at specific resonances ($\mathrm{T}_2$ and $\mathrm{T}_6$ at
$B=5\,\mbox{T}$, $\mathrm{T}_5$ and $\mathrm{T}_6$ at $B=8\,\mbox{T}$) and the local maxima at $\nu=0$ for the
$\mathrm{T}_1$ transitions. Moreover, the local minima of $v_\mathrm{F}^*$ for $\mathrm{T}_2$ at $\nu=\pm4$, when the
$1$th or $-1$th Landau level is half-filled, which occur only at $B=8\,\mbox{T}$ and are absent in other fields, are
not predicted by our approach.

To characterize an accuracy of our fitting, we present in Fig.~\ref{Fig3}(d) the root mean square deviation
\begin{eqnarray}
\Delta=\sqrt{\sum_{i=1}^N \frac{[(v_\mathrm{F}^*)_i^\mathrm{calc}-(v_\mathrm{F}^*)_i^\mathrm{exp}]^2}{N}} \label{rms}
\end{eqnarray}
of calculated renormalized Fermi velocities $(v_\mathrm{F}^*)_i^\mathrm{calc}$ from $N$ experimental values
$(v_\mathrm{F}^*)_i^\mathrm{exp}$ (here it means for all fields and all resonances at once, 260 points in total). As
functions of the fitting parameter $\varepsilon$, $\Delta$ reach rather sharp minima at optimal $\varepsilon$ in each
approximation. The minimal $\Delta$ about $0.015\times10^6\,\mbox{m/s}$ are comparable to the experimental
uncertainties of $(v_\mathrm{F}^*)_i^\mathrm{exp}$ \cite{Russell}, so the fitting can be considered to be sufficiently
accurate.

For completeness of the analysis, we can also consider the intraband transitions $n\rightarrow n+1$ and
$-n-1\rightarrow-n$. Several examples calculated in the conditions of the experiment \cite{Russell} are presented in
Fig.~\ref{Fig4}. Each $n\rightarrow n+1$ ($-n-1\rightarrow-n$) transition exists in the range $4n-2<\nu<4n+6$
($-4n-6<\nu<-4n+2$) of the filling factors, and the transition energies are minimal at $\nu=4n+2$ ($\nu=-4n-2$). These
minima are caused by the excitonic correction (\ref{E_exc}), which is maximal when the initial Landau level is
completely filled, and the final level is completely empty. We can also note that $v_\mathrm{F}^*$ again decreases with
increasing the doping level due to enhancement of the screening, while the Hartree-Fock approximation misses this
effect and greatly overestimates the variations of $v_\mathrm{F}^*$ vs. $\nu$. The electron-hole asymmetry for these
transitions is negligible.

\begin{figure}[b]
\begin{center}
\resizebox{0.8\columnwidth}{!}{\includegraphics{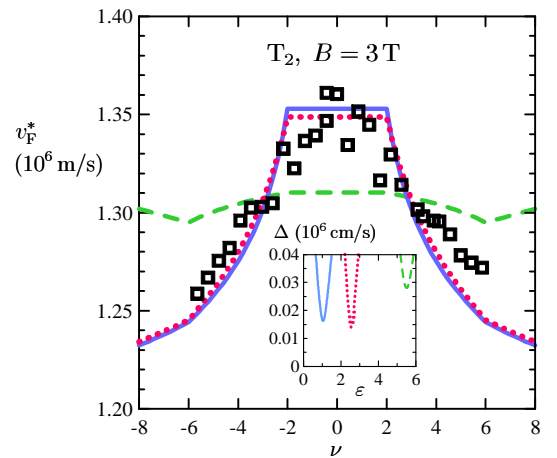}}
\end{center}
\caption{\label{Fig5} Renormalized Fermi velocity $v_\mathrm{F}^*$ for the $T_2$ transition ($-1\rightarrow
2/-2\rightarrow1$) at $B=3\,\mbox{T}$, taken from the experiment \cite{Sonntag} (squares) and calculated theoretically
in the Hartree-Fock approximation (dashed lines), with taking into account the interaction screening (solid lines) and
with the self-consistent screening (dotted lines). The dielectric constants $\varepsilon$, used in each approximation,
are listed in the second line of Table~\ref{Table1}. Inset shows root mean square deviations (\ref{rms}) between the
calculated and experimental $v_\mathrm{F}^*$ as functions of $\varepsilon$ in the three approximations.}
\end{figure}

Another experiment we analyze is Ref.~\cite{Sonntag} where graphene is suspended $160\,\mbox{nm}$ above oxidized
silicon, and the filling-factor dependence of the $\mathrm{T}_2$ transition energy was measured at $B=3\,\mbox{T}$ by
observing its avoided crossing with the phonon energy in Raman spectrum. In Fig.~\ref{Fig5} we plot the results of our
calculations for this transition in the three approximations at optimal $\varepsilon$ listed in the second line of
Table~\ref{Table1}. We observe the same regularities as in the previous case. The Hartree-Fock approximations requires
overestimated $\varepsilon$ and cannot explain the rapid falloff of $v_\mathrm{F}^*$ at $|\nu|>2$. At $|\nu|>6$ we see
even slight increase of $v_\mathrm{F}^*$ due to suppression of the excitonic correction when the $2$nd or $-2$nd Landau
level start to be partially filled. In contrast, with taking into account the screening we obtain the realistic
$\varepsilon$ for graphene suspended above the oxidized silicon, and the falloff is well reproduced. Nevertheless, the
experimental points demonstrate an additional maximum at $\nu=0$. This is not described by our approach, which predicts
plateaus at $|\nu|<2$, as discussed in Sec.~\ref{sec3}.

The root mean square deviations (\ref{rms}), calculated for $31$ experimental points, are shown in the inset to
Fig.~\ref{Fig5} and demonstrate pronounced minima at the optimal $\varepsilon$. The minimal $\Delta$ about
$0.015\times10^6\,\mbox{m/s}$, achieved with the screened interaction, are comparable to the experimental uncertainties
$(0.01\div0.05)\times10^6\,\mbox{m/s}$ \cite{Sonntag}.

\section{Landau level broadening}\label{sec5}
One more experiment where the filling-factor dependent transition energy was measured is Ref.~\cite{Henriksen}. In this
earlier work, graphene layer lied directly on an oxidized silicon substrate and carrier mobility was one-two orders of
magnitude lower than in the aforementioned works \cite{Russell,Sonntag} due to charged impurities in the substrate. The
$\mathrm{T}_1$ cyclotron resonance was studied at $B=18\,\mbox{T}$ and the unusual W-shaped form of the transition
energy vs. $\nu$ was found with the local maximum at $\nu=0$ and two minima at integer Landau level fillings
$\nu=\pm2$.

\begin{figure}[t]
\begin{center}
\resizebox{0.75\columnwidth}{!}{\includegraphics{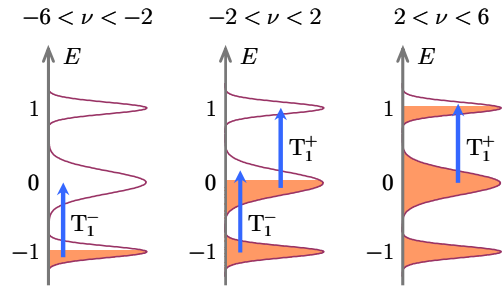}}
\end{center}
\caption{\label{Fig6} Broadened Landau levels $n=0,\pm1$ (not in scale) and cyclotron transitions between them when
these levels are partially filled.}
\end{figure}

To explain these results, we need to take into account disorder, because at mobilities of several thousands of
$\mbox{cm}^2/\mbox{V}\cdot\mbox{s}$, reported in \cite{Henriksen}, the disorder-induced Landau level widths
$\sim20\,\mbox{meV}$ become comparable with the energy scale $e^2/\varepsilon l_H$ of Coulomb interaction effects. The
main mechanism of disorder effect on the transition energies is the following. Assume that Landau levels are broadened
giving rise to Gaussian mini-bands in the density of states, as shown in Fig.~\ref{Fig6}. At partial filling of each
level, its mini-band is partially filled, so the average energy of the filled (empty) electron states is lower (higher)
than the band center where the unperturbed Landau level energy would be located. As a result, the average transition
energy increases due to Landau level broadening in addition to interaction effects when either initial or final level
is partially filled ($\nu\neq\pm2$ in our case). The similar effect was discussed in \cite{Ando75} for a
two-dimensional gas of massive electrons in the framework of self-consistent Born approximation.

To describe this effect, we assume the Gaussian spectral density
$\rho_n(E)=(\sqrt{2\pi}\Gamma_n)^{-1}\exp[-(E-E_n)^2/2\Gamma_n^2]$ for each $n$th partially filled broadened level.
Integrating it up to the Fermi level $\mu$ and assuming low temperature, we find the occupation number $f_n$, and,
using (\ref{f_n}), we get the relationship between $\nu$ and $\mu$: $\nu=4n+2\Phi([\mu-E_n]/\sqrt2\Gamma_n)$, where
$\Phi$ is the error function. The disorder-induced correction $\langle\Delta\Omega_n\rangle$ to the transition energy
is a difference between the average energies (relative to the band centers) of empty states on a final Landau level and
filled states on an initial level. It should be additionally weighted according to (\ref{Omega_av}), when $-2<\nu<2$
and thus both transitions $\mathrm{T}_1^\pm$ are present, resulting in:
\begin{eqnarray}
\langle\Delta\Omega_n\rangle=\left\{\begin{array}{rcl}\displaystyle\sqrt{\frac2\pi}
\frac{\Gamma_{-1}e^{-\frac{(\mu-E_{-1})^2}{2\Gamma_{-1}^2}}}{3+\nu/2},&\mbox{if}&-6<\nu<-2,\\
\displaystyle\sqrt{\frac2\pi}\Gamma_0e^{-\frac{(\mu-E_0)^2}{2\Gamma_0^2}},&\mbox{if}&-2<\nu<2,\\
\displaystyle\sqrt{\frac2\pi}
\frac{\Gamma_1e^{-\frac{(\mu-E_1)^2}{2\Gamma_1^2}}}{3-\nu/2},&\mbox{if}&\hphantom{-}2<\nu<6.
\end{array}\right.\label{Delta_Omega}
\end{eqnarray}
This dependence has a maximum at $\nu=0$ and minima at $\nu=\pm2$ in accordance with the experiment \cite{Henriksen}.

Another effect of the disorder is the presence of interlevel transitions when any $n$th Landau level is partially
filled, which provide an extra contribution to the screening. In the simplest approximation, they lead to the
polarizability of the Thomas-Fermi kind
\begin{eqnarray}
\Pi^\mathrm{TF}_n(q)=-gF_{nn}(q)\rho_n(\mu),\label{Pi_TF}
\end{eqnarray}
which was used in \cite{Yang} to study Landau level broadening in graphene.

We use the self-consistent Born approximation for a polarizability in magnetic field, which was originally developed in
\cite{Ando71,Ando74} for a two-dimensional electron gas with short-range impurities. In our work, we assume the
disorder to be long-ranged, because the main origin of disorder in graphene on a $\mathrm{SiO}_2$ substrate are
long-range charged impurities \cite{DasSarma}. Introducing the mean square $\langle U^2\rangle$ of the slowly varying
disorder potential $U(\mathbf{r})$, we get the following polarizability of disordered graphene (see the similar
formulas in \cite{Ando71,Ando74} obtained by summing an impurity ladder in a polarization loop):
\begin{eqnarray}
\Pi^\mathrm{D}(q,i\omega)=g\sum_{nn'}F_{nn'}(q)\nonumber\\
\times T\sum_\epsilon\frac{G^\mathrm{D}_{n'}(i\epsilon+i\omega)G^\mathrm{D}_n(i\epsilon)} {1-\langle U^2\rangle
G^\mathrm{D}_{n'}(i\epsilon+i\omega)G^\mathrm{D}_n(i\epsilon)},\label{Pi_D1}
\end{eqnarray}
where $G^\mathrm{D}_n(i\epsilon)=\int dE\:\rho_n(E)/(i\omega-E+\mu)$ is the Green function of electron on the $n$th
Landau level in the presence of disorder. Instead of a half-elliptic spectral density \cite{Yang}, which is known to be
an artefact of the self-consistent Born approximation \cite{MacDonald}, we use, as above, the Gaussian function
$\rho_n(E)$.

\begin{figure}[b]
\begin{center}
\resizebox{0.75\columnwidth}{!}{\includegraphics{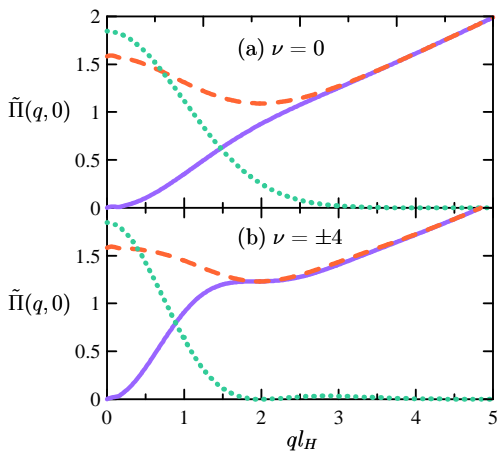}}
\end{center}
\caption{\label{Fig7} Dimensionless static polarizability of graphene in magnetic field $\tilde\Pi(q,0)=-(2\pi
v_\mathrm{F}l_H/g)\Pi(q,0)$, where $l_H=\sqrt{\hbar c/eH}$, calculated (a) when the 0th Landau level is half-filled,
$\nu=0$, (b) when the 1th or $-1$th level is half-filled, $\nu=\pm4$. Solid lines: clean graphene (\ref{Pi}), dashed
lines: disordered graphene (\ref{Pi_D3}), dotted lines: the Thomas-Fermi approximation (\ref{Pi_TF}). Calculation
parameters are $v_\mathrm{F}=0.85\times10^6\,\mbox{m/s}$, $B=18\,\mbox{T}$, $\Gamma_0=\Gamma_{\pm1}=20\,\mbox{meV}$.}
\end{figure}

Taking the static limit $i\omega\rightarrow0$ and switching in (\ref{Pi_D1}) from the frequency summation to an
integration along the branch cut at $\mathrm{Im}(i\epsilon)=0$, we get in the limit $T\rightarrow0$:
\begin{eqnarray}
\Pi^\mathrm{D}(q,0)=-\frac{g}\pi\sum_{nn'}F_{nn'}(q)\nonumber\\
\times\int\limits_{-\infty}^0\mathrm{Im}\frac{G^\mathrm{D}_{n'}(z+i\delta)G^\mathrm{D}_n(z+i\delta)} {1-\langle
U^2\rangle G^\mathrm{D}_{n'}(z+i\delta)G^\mathrm{D}_n(z+i\delta)}.\label{Pi_D2}
\end{eqnarray}
This polarizability consists of two physically distinct parts. The first one is the contribution of interlevel
transitions with $n\neq n'$. It does not differ too much from than in a clean system (\ref{Pi}) if the widths of Landau
levels $\Gamma_n$ are much smaller than interlevel separations. The second one is the contribution of intralayer
transitions $n=n'$ arising when the $n$th layer is partially filled. Taking the disorder strength to be equal to the
Landau level width $\sqrt{\langle U^2\rangle}=\Gamma_n$, as follows from calculations of $G^\mathrm{D}$ with the
long-range disorder, we get the static polarizability of disordered graphene:
\begin{eqnarray}
\Pi^\mathrm{D}(q,0)\approx\Pi(q,0)-gF_{nn}(q)\int\limits_{-\infty}^0\mathrm{Im}\frac{[G^\mathrm{D}_n(z+i\delta)]^2}
{1-\Gamma_n^2 [G^\mathrm{D}_n(z+i\delta)]^2}\label{Pi_D3}
\end{eqnarray}
and use it in the following calculations.

Fig.~\ref{Fig7} shows the examples of static polarizabilities calculated at half-fillings of 0th and $\pm1$th Landau
levels. In a clean graphene, $\Pi(q,0)\propto q^2$ at $q\rightarrow0$ since the system becomes insulating in magnetic
field, and the only source of the screening are gapped interlevel transitions. Disorder makes $\Pi^\mathrm{D}(q,0)$
nonzero at $q\rightarrow0$ due to intralevel transitions. The Thomas-Fermi approximation, by taking into account only
the latter, provides a wrong short-wavelength asymptotic of the polarizability $\Pi^\mathrm{TF}_n(q)$, which should
tend to the polarizability of undoped graphene $\Pi(q,0)=-gq/16\hbar v_\mathrm{F}$ \cite{CastroNeto}.

\begin{figure}[t]
\begin{center}
\resizebox{0.8\columnwidth}{!}{\includegraphics{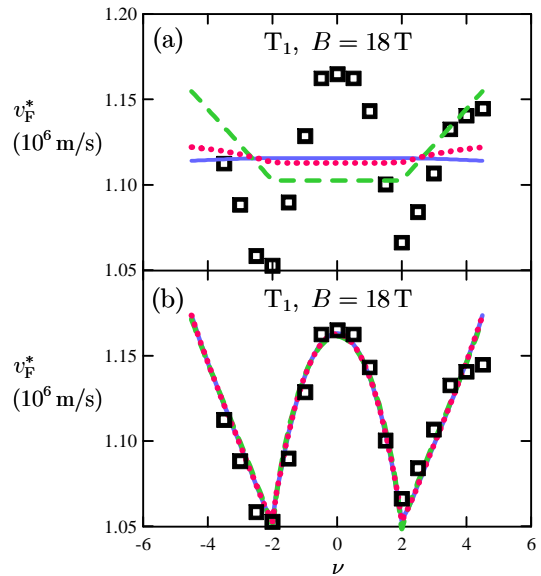}}
\end{center}
\caption{\label{Fig8} Renormalized Fermi velocity $v_\mathrm{F}^*$ for the $T_1$ transition ($0\rightarrow
1/-1\rightarrow0$) at $B=18\,\mbox{T}$, taken from the experiment \cite{Henriksen} (squares) and calculated
theoretically for (a) clean and (b) disordered graphene. The calculations are carried out in the Hartree-Fock
approximation (dashed lines), with taking into account the interaction screening (solid lines) and with the
self-consistent screening (dotted lines). The dielectric constants $\varepsilon$ used in each calculation are listed in
Table~\ref{Table2}.}
\end{figure}

We calculated the renormalized Fermi velocity, corresponding to the weighted energy (\ref{Omega_av}) of the $T_1$
transition with taking into account the correction (\ref{Delta_Omega}) and the screening (\ref{Pi_D3}) in the
disordered system. For comparison, we carried out the same calculations for the clean system, as did in the previous
section. The results of the fitting of experimental points from Ref.~\cite{Henriksen} are shown in Fig.~\ref{Fig8}, and
the calculation parameters are listed in Table~\ref{Table2}.

\begin{table}[b]
\caption{\label{Table2}First two lines: dielectric constants of surrounding medium $\varepsilon$, which provide the
best least-square fittings of the experimental data from Ref.~\cite{Henriksen} at
$v_\mathrm{F}=0.85\times10^6\,\mbox{m/s}$ in the three approximations for the interaction listed in Sec.~\ref{sec4} for
clean or disordered system. For disordered system, the widths of $0$th and $\pm1$st Landau levels are also specified in
the last two lines.}
\centering
\begin{tabular}{ccccc}
\hline%
\hline%
&Unscreened&Screened&Self-consistent\\
System&interaction&interaction&screening\\
\hline%
Clean&7.26&2.82&3.85\\
\hline
Disordered&9.24&4.95&5.74\\
$\Gamma_0\,\mbox{(meV)}$&22&25&23\\
$\Gamma_{\pm1}\,\mbox{(meV)}$&12&19&17\\
\hline%
\hline%
\end{tabular}
\end{table}

For the values of $\varepsilon$, we observe the same regularities as noted in the previous section. These values are
close to those obtained in our earlier analysis \cite{Sokolik1} of cyclotron resonance data for graphene on
$\mathrm{SiO}_2$. However the most drastic effects come from inclusion of disorder: while in the clean system
$v_\mathrm{F}^*$ has the plateau at $|\nu|<2$ and remain the same (or slightly increases due to suppression of the
excitonic correction) at $|\nu|>2$, in the disordered system it has the parabolic-like maximum at $\nu=0$ and the sharp
minima at $\nu=\pm2$, just as the experiment shows. The values of Landau level widths $\Gamma_n$ obtained via the
fitting procedure ($15-25\,\mbox{meV}$) look realistic, since they are close to typical widths of spectral lines
observed in the same experiment \cite{Henriksen} and in other works on graphene on a $\mathrm{SiO}_2$ substrate
\cite{Luican}. The minimal value of the root mean square deviation (\ref{rms}) is about $0.009\times10^6\,\mbox{m/s}$
in this case.

\section{Conclusions}\label{sec6}
We present detailed calculations of the inter-Landau level cyclotron transition energies in graphene in strong magnetic
fields taking into account Coulomb interaction between massless Dirac electrons. Calculating the optical conductivity
and solving the vertex equation in the static random-phase approximation with the excitonic ladder, we found the
many-body corrections to the transition energies coming from the self-energy and excitonic effects. We show that the
cyclotron transition lines can be split in doped graphene for opposite circular polarizations because of the
electron-hole asymmetry of exchange self-energies, although this splitting may be unobservable if these lines are
sufficiently wide or either a linearly polarized or unpolarized light is used. By this reason, we calculate the
weighted transition energy for both polarizations at once and convert it to the renormalized Fermi velocity
$v_\mathrm{F}^*$ for each transition.

The dependence of $v_\mathrm{F}^*$ on the Landau level filling factor $\nu$ is analyzed. In the mean-field
approximation, $v_\mathrm{F}^*(\nu)$ has a plateau at $-2<\nu<2$ due to a partial cancelation of the self-energy and
excitonic effects and rapidly decreases at $|\nu|>2$ due to enhancement of the screening. Our calculations, carried out
with the bare Fermi velocity $v_\mathrm{F}=0.85\times10^6\,\mbox{m/s}$ and with the dielectric constant of surroundings
$\varepsilon$, treated as an adjustable parameter, showed good agreement with two recent experiments
\cite{Russell,Sonntag} on high-mobility graphene samples, when the screening by graphene electrons is taken into
account. The obtained phenomenological $\varepsilon$ describe the external dielectric screening not only by an
underlaying substrate, but also by adjacent hexagonal boron nitride layers. The Hartree-Fock approximation, which
neglects the density-dependent screening by graphene electrons, fails to explain the observed rapid decrease of
$v_\mathrm{F}^*$ at $|\nu|>2$.

Our calculations for the intraband transitions $n\rightarrow n+1$ and $-n-1\rightarrow-n$ predict the V-like dependence
$v_\mathrm{F}^*(\nu)$ with the minima at, respectively, $\nu=4n+2$ and $\nu=-4n-2$ caused by the excitonic effects.
Existence of these minima can be verified experimentally, although an accurate detection of the interband transition
lines can be challenging (but possible \cite{Sadowski1}) due to their much lower energies: even for the the highest
magnetic fields 20-30 T these energies are below 100 meV.

We also describe the data of the earlier cyclotron resonance experiment \cite{Henriksen} with graphene sample on
$\mathrm{SiO}_2$, where carrier mobility is much lower. In this case we take into account long-range disorder, which
broadens Landau levels and thus shifts the resonant energy upward when initial or final level is partially filled, and
induces the interlevel transitions contributing to the screening. Assuming the Gaussian spectral density for the $0$th
and $\pm1$th broadened Landau levels, we achieved good agreement with the experiment and explained the main features of
the $v_\mathrm{F}^*(\nu)$ dependence: the parabolic-like maximum at $\nu=0$ and the sharp minima at $\nu=\pm2$.

As shown, the combined action of exchange interaction, excitonic effects, interaction screening and disorder should be
taken into account when considering graphene in strong magnetic field. Our approach takes into consideration these
factors and thus allowed us to explain main features of the filling-factor dependent experimental data
\cite{Russell,Sonntag,Henriksen}, which would be hardly possible within the Hartree-Fock approximation
\cite{Iyengar,Bychkov,Barlas,Roldan1,Shizuya1,Shizuya2,Faugeras1,Chizhova,Shizuya4,Shizuya2018} where the screening and
Landau level broadening are neglected. However, some issues remain to be clarified. In particular, the mean-field
approach does not describe the $\Lambda$-shaped maxima of $v_\mathrm{F}^*$ at $\nu=0$ observed in
\cite{Russell,Sonntag} for $\mathrm{T}_1$ transitions, the minima at $\nu=\pm4$ observed for $\mathrm{T}_2$ at
$B=8\,\mbox{T}$ in \cite{Russell}, and a possible splitting of the $\mathrm{T}_1$ transition line observed in
\cite{Russell}. All these features go beyond the mean-field theory for massless Dirac electrons and can be attributed
to some unaccounted role of disorder, finite size effects, Moire superlattice potential from adjacent boron nitride
layers \cite{ChenNComm}, Landau level splitting \cite{Goerbig,Song} or electron dynamics on a partially filled level
\cite{Basko}. Note that assumption of a substrate-induced band gap allowed to explain some features of the experimental
data of \cite{Russell} in the recent work \cite{Shizuya2018}, so a further analysis in this direction with considering
possible symmetry breakings and gap formation in a system of Dirac electrons together with the interaction, screening
and disorder seems to be promising.

\section*{Acknowledgments}
The work was supported by the grant No. 17-12-01393 of the Russian Science Foundation.

\appendix

\begin{widetext}
\section{Electron wave functions}\label{Appendix_A}
Similarly to \cite{Pyatkovskiy,Roldan3,Lozovik,Goerbig}, we describe single-particle states of massless electrons in
magnetic field $\mathbf{H}$ using the symmetric gauge $\mathbf{A}=\frac12\left[\mathbf{H}\times\mathbf{r}\right]$. In
the absence of a valley splitting or intervalley transitions, it is sufficient to consider the electrons only in the
$\mathbf{K}$ valley, where the Dirac Hamiltonian is
\begin{eqnarray}
H=v_\mathrm{F}\left(\mathbf{p}-\frac{e}c\mathbf{A}\right)\cdot\boldsymbol\sigma=\frac{\hbar
v_\mathrm{F}\sqrt2}{l_H}\left(\begin{array}{cc}0&a\\a^+&0\end{array}\right).\label{H_0}
\end{eqnarray}
Here $l_H=\sqrt{\hbar c/|e|H}$ is the magnetic length (we assume $e<0$ in this section), $a=l_Hp_-/\hbar-ir_-/2l_H$ and
$a^+=l_Hp_+/\hbar+ir_+/2l_H$ are, respectively, lowering and raising operators obeying the commutation relation
$[a,a^+]=1$, and $p_\pm=(p_x\pm ip_y)/\sqrt2$, $r_\pm=(x\pm iy)/\sqrt2$.

Introducing the complementary set of ladder operators \cite{Goerbig} $b=l_Hp_+/\hbar-ir_+/2l_H$,
$b^+=l_Hp_-/\hbar+ir_-/2l_H$, which obey $[b,b^+]=1$ and commute with $a$, $a^+$, we can construct the states of a
two-dimensional oscillator $|\phi_{nk}\rangle=(a^+)^n(b^+)^k|\phi_{00}\rangle/\sqrt{n!k!}$ with the wave functions in
polar coordinates:
\begin{eqnarray}
\phi_{nk}(r,\varphi)=\frac{i^{|n-k|}}{\sqrt{2\pi}l_H}\sqrt{\frac{\min(n,k)!}{\max(n,k)!}}\:e^{-r^2/4l_H^2}
\left(\frac{r}{\sqrt2l_H}\right)^{|n-k|}e^{i(n-k)\varphi}L_{\min(n,k)}^{|n-k|}\left(\frac{r^2}{2l_H^2}\right),
\label{phi_nk}
\end{eqnarray}
where $L_n^m(x)$ are the associated Laguerre polynomials. The eigenfunctions of the Hamiltonian (\ref{H_0}) are
\cite{Gumbs,Roldan2,Roldan3,Barlas,Iyengar,Roldan1,Goerbig,Bychkov,Lozovik}
\begin{eqnarray}
\psi_{nk}=(\sqrt{2})^{\delta_{n0}-1}\left(\begin{array}{r}s_n\phi_{|n|-1,k}\\
\phi_{|n|,k}\end{array}\right),\label{psi_nk}
\end{eqnarray}
and eigenvalues are (\ref{E_n}). Here $n=0,\pm1,\pm2,\ldots$ is the Landau level number, $k=0,1,2,\ldots$ is the
guiding center index responsible for Landau levels degeneracy, $s_n\equiv\mathrm{sign}\,(n)$, and we assume that
$\phi_{nk}=0$ if $n$ or $k$ is negative.

The bare electron Green function in the Matsubara representation $G(\mathbf{r},\mathbf{r}',\tau)=-\langle
T_\tau\Psi(\mathbf{r},\tau)\Psi^+(\mathbf{r}',0)\rangle$ can be constructed from (\ref{psi_nk}):
\begin{eqnarray}
G_0(\mathbf{r},\mathbf{r}',i\epsilon)=\sum_{nk}\frac{\psi_{nk}(\mathbf{r})\psi_{nk}^+(\mathbf{r}')}{i\epsilon-E_n+\mu},
\label{G0}
\end{eqnarray}
where $\mu$ is the chemical potential; note $G_0$ is the $(2\times2)$ matrix in the sublattice space.

Using the table integral Eq. 2.20.16.10 from \cite{Prudnikov}, we can present (\ref{phi_nk}) in Cartesian coordinates
as
\begin{eqnarray}
\phi_{nk}(x,y)=\frac{i^{n-k}}{\sqrt{2\pi}}\int_{-\infty}^{+\infty}dt\:e^{ity} \varphi_n\left(l_Ht+x/2l_H\right)
\varphi_k\left(l_Ht-x/2l_H\right),\label{phi_nk_int}
\end{eqnarray}
where $\varphi_n(x)=e^{-x^2/2}H_n(x)/\sqrt{2^nn!\sqrt\pi}$ are the dimensionless eigenfunctions of quantum
one-dimensional harmonic oscillator, $H_n(x)$ are Hermite polynomials. Then, using (\ref{phi_nk_int}), and
orthonormality and completeness of the basis $\{\varphi_n(x)\}$, a lot of useful transformation rules for $\phi_{nk}$
can be obtained, for example, the summation formula
\begin{eqnarray}
\sum_{k=0}^\infty\phi_{n_1k}(\mathbf{r_1})\phi^*_{n_2k}(\mathbf{r}_2)=
\frac{e^{i(\mathbf{r}_1\mathbf{r}_2\mathbf{e}_z)/2l_H^2}}{\sqrt{2\pi}l_H}\,
\phi_{n_1n_2}(\mathbf{r}_1-\mathbf{r}_2)\label{k_sum}
\end{eqnarray}
and the form-factor of Landau level wave functions (see also \cite{Pyatkovskiy})
\begin{eqnarray}
\int d\mathbf{r}\:e^{i\mathbf{q}\mathbf{r}}\phi^*_{n_1n_2}(\mathbf{r})\phi_{n_3n_4}(\mathbf{r})=2\pi
l_H^2\phi^*_{n_1n_3}(\mathbf{a}_\mathbf{q})\phi_{n_2n_4}(\mathbf{a}_\mathbf{q}),
\qquad\mathbf{a}_\mathbf{q}\equiv-l_H^2[\mathbf{e}_z\times\mathbf{q}].\label{formfactor}
\end{eqnarray}

\section{Exchange self-energies}\label{Appendix_B}
Exchange self-energy acquired by an electron in the state $\psi_{nk}$ is given by the usual Fock expression
\begin{eqnarray}
\Sigma_{nk}^\mathrm{exch}=-\sum_{n'k'}f_{n'k'}\int d\mathbf{r}_1d\mathbf{r}_2\:v(\mathbf{r}_1-\mathbf{r}_2)
\psi^+_{nk}(\mathbf{r}_1)\psi_{n'k'}(\mathbf{r}_1) \psi^+_{n'k'}(\mathbf{r}_2)\psi_{nk}(\mathbf{r}_2).
\end{eqnarray}
After the Fourier transform of the Coulomb interaction $v(\mathbf{r})=(2\pi)^{-2}\int d\mathbf{q}\:v_q
e^{i\mathbf{qr}}$ and using (\ref{k_sum})--(\ref{formfactor}), we get
\begin{eqnarray}
\Sigma_{nk}^\mathrm{exch}=-\sum_{n'}f_{n'}\langle nn'|v|n'n\rangle
\end{eqnarray}
with the exchange matrix elements of Coulomb interaction defined as
\begin{eqnarray}
\langle nn'|v|n'n\rangle=2^{\delta_{n0}+\delta_{n'0}-2}\frac{l_H^2}{2\pi}\int
d\mathbf{q}\:v_q\left|s_ns_{n'}\phi_{|n|-1,|n'|-1}(\mathbf{a}_\mathbf{q})
+\phi_{|n||n'|}(\mathbf{a}_\mathbf{q})\right|^2.
\end{eqnarray}
We assumed that the occupation numbers do not depend on $k'$, $f_{n'k'}\equiv f_{n'}$, and the resulting
$\Sigma_{nk}^\mathrm{exch}$ turns out to be also independent on $k$, so the Landau level degeneracy is preserved. By
replacing $v_q$ with the statically screened interaction $V(q)$, as depicted in Fig.~\ref{Fig1}(c), we get the screened
exchange energy (\ref{Sigma_exch})--(\ref{exch_me}), and the bare electron Green function (\ref{G0}) becomes
``dressed'' with the interaction and turns into
\begin{eqnarray}
G(\mathbf{r},\mathbf{r}',i\epsilon)=
\sum_{nk}\frac{\psi_{nk}(\mathbf{r})\psi_{nk}^+(\mathbf{r}')}{i\epsilon-E_n-\Sigma_n+\mu}.\label{G}
\end{eqnarray}

\section{Vertex equation}\label{Appendix_C}
Introducing the Green function for currents $G^j_{\alpha\beta}(\mathbf{r},\mathbf{r}',\tau)=-\langle
T_\tau\Psi^+(\mathbf{r},\tau)\sigma_\alpha\Psi(\mathbf{r},\tau)
\Psi^+(\mathbf{r}',0)\sigma_\beta\Psi(\mathbf{r}',0)\rangle$, we can write the conductivity (\ref{sigma1}) as
\begin{eqnarray}
\sigma_{\alpha\beta}(\mathbf{q},\omega)=\frac{ie^2v_\mathrm{F}^2}{\hbar\omega S}\int
d\mathbf{r}d\mathbf{r}'\:e^{-i\mathbf{q}(\mathbf{r}-\mathbf{r}')}
G^j_{\alpha\beta}(\mathbf{r},\mathbf{r}',\hbar\omega+i\delta), \label{sigma2}
\end{eqnarray}
where $G^j_{\alpha\beta}$ can be calculated, as shown in Fig.~\ref{Fig1}(a), from the $(2\times2)$ vertex matrix:
\begin{eqnarray}
G^j_{\alpha\beta}(\mathbf{r},\mathbf{r}',i\omega)=T\sum_{\varepsilon}\int
d\mathbf{r}_1d\mathbf{r}_2\:\mathrm{Tr}\left[\sigma_\alpha G(\mathbf{r},\mathbf{r}_1,i\epsilon+i\omega)
\Gamma_\beta(\mathbf{r}_1,\mathbf{r}_2,\mathbf{r}',i\epsilon,i\omega)
G(\mathbf{r}_2,\mathbf{r},i\epsilon)\right].\label{Gj}
\end{eqnarray}
Here the sum is taken over the fermionic Matsubara frequencies $\epsilon=\pi T(2n+1)$.

The vertex equation in the mean-field (or ladder) approximation, depicted in Fig.~\ref{Fig1}(b), is written
analytically as
\begin{eqnarray}
\Gamma_\beta(\mathbf{r}_1,\mathbf{r}_2,\mathbf{r}',i\epsilon,i\omega)=
\delta(\mathbf{r}_1-\mathbf{r}')\delta(\mathbf{r}_2-\mathbf{r}')\sigma_\beta-T\sum_{\epsilon'}\int
d\mathbf{r}_1'd\mathbf{r}_2'\:V(\mathbf{r}_1-\mathbf{r}_2,i\epsilon-i\epsilon')\nonumber\\ \times
G(\mathbf{r}_1,\mathbf{r}_1',i\epsilon'+i\omega)\Gamma_\beta(\mathbf{r}_1',\mathbf{r}_2',\mathbf{r}',i\epsilon',i\omega)
G(\mathbf{r}_2',\mathbf{r}_2,i\epsilon').\label{vertex1}
\end{eqnarray}
To solve it, we can use the basis of magnetoexcitonic states of Dirac electrons in the symmetric gauge, which were
described earlier in \cite{Lozovik} in slightly different notation:
\begin{eqnarray}
\Psi_{\mathbf{P}n_1n_2}(\mathbf{r}_1,\mathbf{r}_2)=\frac1{2\pi}
e^{i\mathbf{P}(\mathbf{r}_1+\mathbf{r}_2)/2+i(\mathbf{r}_1\mathbf{r}_2\mathbf{e}_z)/2l_H^2}
\Phi_{n_1n_2}(\mathbf{r}_1-\mathbf{r}_2-\mathbf{a}_\mathbf{P}).\label{Psi_me}
\end{eqnarray}
Here $\mathbf{P}$ is the conserved magnetic momentum of the electron-hole pair and
\begin{eqnarray}
\Phi_{n_1n_2}(\mathbf{r})=\sqrt2^{\delta_{n_10}+\delta_{n_20}-2}
\left(\begin{array}{cc}s_{n_1}s_{n_2}\phi_{|n_1|-1,|n_2|-1}(\mathbf{r})&
s_{n_1}\phi_{|n_1|-1,|n_2|}(\mathbf{r})\\s_{n_2}\phi_{|n_1|,|n_2|-1}(\mathbf{r})&
\phi_{|n_1|,|n_2|}(\mathbf{r})\end{array}\right)\label{Phi_nk}
\end{eqnarray}
is the matrix wave function of relative motion of electron and hole written in the basis of their sublattices $A$, $B$.
Using (\ref{phi_nk_int}), the unitary transformations between the magnetoexcitonic states and the states (\ref{psi_nk})
of individual electron and hole can be derived:
\begin{eqnarray}
\psi_{n_1k_1}(\mathbf{r}_1)\psi^+_{n_2k_2}(\mathbf{r}_2)=l_H^2\!\!\int
d\mathbf{P}\:\phi_{k_1k_2}^*(\mathbf{a}_\mathbf{P}) \Psi_{\mathbf{P}n_1n_2}(\mathbf{r}_1,\mathbf{r}_2),\quad
\Psi_{\mathbf{P}n_1n_2}(\mathbf{r}_1,\mathbf{r}_2)=l_H^2\!\!\sum_{k_1k_2}\phi_{k_1k_2}(\mathbf{a}_\mathbf{P})
\psi_{n_1k_1}(\mathbf{r}_1)\psi^+_{n_2k_2}(\mathbf{r}_2).\label{transform}
\end{eqnarray}

Projecting the vertex matrix $\Gamma_\beta$ on the magnetoexcitonic states
\begin{eqnarray}
\Gamma_{\beta,\mathbf{P}n_1n_2}(\mathbf{r}',i\epsilon,i\omega)=\int d\mathbf{r}_1d\mathbf{r}_2\:
\mathrm{Tr}\left[\Psi^+_{\mathbf{P}n_1n_2}(\mathbf{r}_1,\mathbf{r}_2)
\Gamma_\beta(\mathbf{r}_1,\mathbf{r}_2,\mathbf{r}',i\epsilon,i\omega)\right]\label{vertex2}
\end{eqnarray}
and using (\ref{G}), (\ref{transform}), we get (\ref{vertex1}) in the electron-hole pair (or magnetoexcitonic)
representation:
\begin{eqnarray}
\Gamma_{\beta,\mathbf{P}n_1n_2}(\mathbf{r}',i\epsilon,i\omega)= \Gamma^{(0)}_{\beta,\mathbf{P}n_1n_2}(\mathbf{r}')
-T\sum_{\epsilon'n_1'n_2'}\int
d\mathbf{P}'\:\frac{\langle\Psi_{\mathbf{P}n_1n_2}|V(i\epsilon-i\epsilon')|\Psi_{\mathbf{P}'n_1'n_2'}\rangle
\Gamma_{\beta,\mathbf{P}'n_1'n_2'}(\mathbf{r}',i\epsilon',i\omega)}
{(i\epsilon'+i\omega-E_{n_1'}-\Sigma_{n_1'}+\mu)(i\epsilon'-E_{n_2'}-\Sigma_{n_2'}+\mu)}.\label{vertex3}
\end{eqnarray}
Here the bare vertex is $\Gamma^{(0)}_{\beta,\mathbf{P}n_1n_2}(\mathbf{r}')=(e^{-i\mathbf{Pr}'}/2\pi)\,\mathrm{Tr}
\left[\Phi^+_{n_1n_2}(\mathbf{a}_\mathbf{P})\sigma_\beta\right]$.

To solve Eq.~(\ref{vertex3}), we use the static approximation $V(\mathbf{r},i\epsilon-i\epsilon')=V(\mathbf{r})$ and
neglect the mixing of different electron-hole pairs in the ladder diagrams, assuming $n_1'=n_1$, $n_2'=n_2$. Therefore
the vertex matrix turns out to be independent on a relative energy of electron and hole $\epsilon$:
\begin{eqnarray}
\Gamma_{\beta,\mathbf{P}n_1n_2}(\mathbf{r}',i\omega)=\frac{e^{-i\mathbf{Pr}'}}{2\pi}\,\mathrm{Tr}
\left[\Phi^+_{n_1n_2}(\mathbf{a}_\mathbf{P})\sigma_\beta\right]\left\{1+\langle n_1n_2|V_\mathbf{P}|n_1n_2\rangle
\frac{f_{n_2}-f_{n_1}}{i\omega+E_{n_2}+\Sigma_{n_2}-E_{n_1}-\Sigma_{n_1}}\right\}^{-1}.\label{vertex4}
\end{eqnarray}
The average interaction energies of magnetoexcitons are $\langle n_1n_2|V_\mathbf{P}|n_1n_2\rangle=\int
d\mathbf{r}\:V(\mathbf{r}-\mathbf{a}_\mathbf{P})
\mathrm{Tr}\left[\Phi^+_{n_1n_2}(\mathbf{r})\Phi_{n_1n_2}(\mathbf{r})\right]$, their counterparts in usual 2D electron
gas were extensively studied earlier \cite{Butov}. Making the Fourier transform $V(\mathbf{r})=(2\pi)^{-2}\int
d\mathbf{q}\:V(q) e^{i\mathbf{qr}}$ and using (\ref{formfactor}), we obtain
\begin{eqnarray}
\langle n_1n_2|V_\mathbf{P}|n_1n_2\rangle=\frac{l_H^2}{2\pi}\int d\mathbf{q}\:V(q)e^{-i\mathbf{qa}_\mathbf{P}}
\mathrm{Tr}\left[\Phi^+_{n_1n_1}(\mathbf{a}_\mathbf{q})\right]
\mathrm{Tr}\left[\Phi_{n_2n_2}(\mathbf{a}_\mathbf{q})\right].\label{exc_me_P}
\end{eqnarray}

The Green function for currents (\ref{Gj}) can by found using (\ref{G}), (\ref{Psi_me}), (\ref{transform}),
(\ref{vertex2}), and (\ref{vertex4}):
\begin{eqnarray}
G_{\alpha\beta}^j(\mathbf{r},\mathbf{r}',i\omega)=\sum_{n_1n_2}\int\frac{d\mathbf{P}}{(2\pi)^2}\,
e^{i\mathbf{P}(\mathbf{r}-\mathbf{r}')}\frac{\mathrm{Tr}\left[\Phi_{n_1n_2}(\mathbf{a}_\mathbf{P})\sigma_\alpha\right]
\mathrm{Tr}\left[\Phi^+_{n_1n_2}(\mathbf{a}_\mathbf{P})\sigma_\beta\right](f_{n_2}-f_{n_1})}
{i\omega+E_{n_2}+\Sigma_{n_2}-E_{n_1}-\Sigma_{n_1}+(f_{n_2}-f_{n_1})\langle n_1n_2|V_\mathbf{P}|n_1n_2\rangle}.
\end{eqnarray}
Substituting it in (\ref{sigma2}) and taking $\mathbf{P}=0$ for optical transitions in (\ref{exc_me_P}), we finally
obtain (\ref{sigma3})--(\ref{exc_me}).

\end{widetext}

\end{document}